\documentclass{sig-alternate-05-2015}
\makeatletter
\def\@copyrightspace{\relax}
\makeatother
\usepackage[utf8]{inputenc}
\usepackage{amsmath}

\usepackage{amsthm}
\usepackage{graphicx}
\usepackage{epstopdf}
\usepackage{amssymb}
\usepackage{amsfonts}
\usepackage{graphicx}
\usepackage{subcaption}
\usepackage{multirow}
\usepackage{bussproofs}
\usepackage[T1]{fontenc}
\usepackage{yfonts}
\usepackage{epstopdf}
\usepackage{comment}
\usepackage[table,dvipsnames]{xcolor}
\usepackage{mathtools}
\usepackage{booktabs}
\usepackage{array}
\usepackage{threeparttable}
\usepackage{cite}
\usepackage{lipsum}
\usepackage{graphicx} 
\graphicspath{ {figures/} }
\usepackage{dcolumn}
\usepackage[rightcaption]{sidecap}
\usepackage{wrapfig}
\usepackage[font={small}]{caption}
\usepackage[shortlabels]{enumitem}

\usepackage{authblk}

\title{Automatic Conversion from Flip-flop to 3-phase Latch-based Designs}

\author{Huimei Cheng}
\author{Yichen Gu}
\author{Peter A. Beerel\thanks{\small This work was partially supported by NSF Grant \#1619415 and DARPA Contract \#HR001119C0070. P. A. Beerel also consults for Galois, Inc. in the area of asynchronous design. }}
\affil{Ming Hsieh Department of Electrical and Computer Engineering, University of Southern California}
\affil{\textit {\{huimeich,yichengu,pabeerel\}@xyz.edu}}

\begin{document}
\maketitle

\begin{abstract}

Latch-based designs have many benefits over their flip-flop based counterparts but have limited use partially because most RTL specifications are flop-centric and automatic conversion of FF to latch-based designs is challenging.
Conventional conversion algorithms target master-slave latch-based designs with two non-overlapping clocks.
This paper presents a novel automated design flow that converts flip-flop to 3-phase latch-based designs. 
The resulting circuits have the same performance as the master-slave based designs but require significantly less latches.
Our experimental results demonstrate the potential for savings in the number of latches (21.3\%), area (5.8\%), and power (16.3\%) on a variety of ISCAS, CEP, and CPU benchmark circuits, compared to the master-slave conversions.

\end{abstract}

\section{Introduction}
\label{sec:introduction}

The growing use of portable/wireless electronic systems and Internet-of-Things (IoT) applications motivates the desire of smaller and more energy-efficient designs in today's very large scale integration (VLSI) circuits.
One of two devices: edge-triggered flip-flops (FFs) or level-sensitive latches are typically used as synchronization and state storage.
It is well-known that latch-based designs can lead to lower power and area than FF-based designs due to time borrowing, smaller cell area, and lower capacitance \cite{Haring2005,singh2018low,Pons2016}, particularly when process variation is considered \cite{Hurst2006}. They are also critical for architecturally-agnostic timing resilient designs \cite{fojtik2012bubble,hand2015blade} which can remove unnecessary margins associated with PVT variations and make near-threshold computing more practical.

As an intermediate between latch and flip-flop based designs, pulsed-latch schemes have also been proposed \cite{lin2014low,paik2010statistical}. These rely on
an edge-triggered pulse generator to provide a short transparency window to all latches. To minimize energy overhead, multi-bit pulsed-latch schemes have been proposed that share pulse generators among several latch cells \cite{singh2018multi}. 
Pulsed-latches, however, must be used carefully because they are subject to hold problems and pulse width variations that are challenging to predict, control, and mitigate (see e.g., \cite{Ding2017}).

A basic challenge to adopting any form of latch-based design is that most RTL specifications are designed using edge 
sensitive FFs.
Approaches to automatically converting an FF- to latch-based design are thus attractive. 
Most conversion flows convert the FF-based designs into pulsed-latch designs \cite{shin2011pulsed} or two-phase latch-based designs controlled by either master-slave clocks \cite{yoshikawa2004timing} or bundled-data asynchronous controllers  \cite{cortadella2006desynchronization,branover2004asynchronous,Saifhashemi2014,zhang2018edge,hand2015blade}.

Optimization of latch-based designs has also been given some attention in the literature. For example, \cite{singh2018low} explores using a mix of master-slave latches and FFs/pulsed-latches. 
Others take advantage of the time borrowing to boost performance and/or reduce area and power consumption \cite{yoshikawa2004timing,singh2018low}. Moreover, retiming algorithms of timing-resilient latch-based designs have been developed that consider not only the number of latches required but also the impact of the amount of needed error-detecting logic \cite{cheng2018automatic}.

Whereas two-phase designs are inherently more robust than pulsed-latch designs, we argue they can be overly restrictive and that
multi-phase latch-based designs \cite{sakallah1990optimal} 
can sometimes be an attractive alternative.

The key contribution of this paper is to demonstrate that a FF-based design can be automatically converted into a robust multi-phase design with fewer latches than a two-phase design. 
In particular, we convert a FF-based to 3-phase latch-based design using a novel Integer Linear Program (ILP) that minimizes latches and retiming to ensure no performance loss.
Our experimental results show an overall average reduction in number of latches of 23\% compared to the conventional master-slave designs on ISCAS89 circuits \cite{ISCAS}, CEP submodules \cite{CEP}, and three CPU designs (i.e. a 3-stage MIPS CPU Plasma \cite{plasma}, a RISC-V Rocket Core \cite{rocket}, and an ARM Cortex-M0 core \cite{armm0}).

This paper is organized as follows. Section \ref{sec:background} introduces background on multi-phase latch-based designs.
Section \ref{sec:LBdesigns} describes the design constraints 
we adopt in our conversion algorithm and the area-performance 
tradeoffs they represent. 
Section \ref{sec:conversion_alg} introduces our ILP-baed conversion algorithm and
Section \ref{sec:results} presents the experimental results based on a broad range of designs.
Finally, some conclusions are drawn in Section \ref{sec:conclusion}.

\section{Background}
\label{sec:background}

The Sakallah, Mudge, and Okulotun (SMO) model \cite{sakallah1990optimal} defines an optimal framework for multi-phase latch-based designs. 
It defines a $k$-phase clock as a collection of $k$ periodic signals with a common cycle time and associated timing constraints, called the General System Timing Constraints (GSTC).
The phases ($p_1$, $p_2$, ... $p_k$) are ordered in a global time reference: $e_{i-1} \leq e_{i} ; \; e_k = T_c $, where $e_i$ is the closing time of phase $p_i$.
$E_{ij}$ is the forward phase shift from phase $p_i$ to phase $p_j$ defined below.
\begin{equation}
\label{eqn:phase_shift}
    E_{ij} = \left \{
    \begin{aligned}
        (e_j - e_i), \; \; & i<j \\
        (T_c + e_j - e_i), \; \; & i \geq j
    \end{aligned}
    \right. 
\end{equation}

Then, the worst-case setup and hold constraints 
for each phase is defined as follows.
\begin{equation}
\label{eqn:GSTC}
\begin{split}
    &\text{Hold:} \;\; H_i \leq d_j + \delta_j +\delta_{ji} - E_{p_j p_i} \\
    &\text{Setup:} \;\; T_c - S_i \geq  D_j + \Delta_j+\Delta_{ji} - E_{p_j p_i}
\end{split}
\end{equation}
Here, $H_i$ and $S_i$ stands for the hold and setup time of the $i^{th}$ latch. 
The shortest (longest) path delay from the $j^{th}$ latch to the $i^{th}$ latch is denoted as $\delta_{ji}$ ($\Delta_{ji}$) and the minimal (maximal) delay value of the $j^{th}$ latch is $\delta_{j}$ ($\Delta_{j}$). 
$d_j$ ($D_j$) represents the earliest (latest) signal departure time, i.e., the amount of 
time after the last $e_j$ that the next data starts to propagate through the $j^{th}$ latch \cite{sakallah1990optimal}. 
$T_c$ denotes the cycle time and we assume all clock phases share the same high pulse width $T_p$ in this paper. 


\section{Latch-Based Designs} 
\label{sec:LBdesigns}

This paper's goal is to convert an FF-based 
to latch-based design minimizing the number of latches based on a reasonable set of constraints. This section explores the implicit trade-offs associated with these constraints and motivates our three-phase clocking approach.

\subsection{Minimal Constraints}

There are two constraints we adopt that are designed to make the application of latch-based designs easier. 
\begin{enumerate}[C1:]
     \item the original position of all FFs must be latched;
     \item neighboring latches, connected by combinational logic, must not be simultaneously transparent;
\end{enumerate}

Constraint C1 is designed to make logical equivalence checking between the latch and FF-designs easier. In particular, we will convert every FF to a 
latch and only add extra latches where necessary to meet these constraints. 
During logical equivalence checking the fixed latches can be viewed as FFs and the extra latches can be treated as transparent. 
Ensuring latches are present at the same position as the original FFs also guarantees the ability to reset the circuit in the same state \cite{singhal1997case}. 

Constraint C2 is designed to avoid min-delay problems. 
In particular,  even with min delay paths equal to 0 ($\delta_i = \delta_{ij} = 0$) the hold constraint is satisfied with zero hold times ($H_i = 0$).\footnote{This follows because constraint C2 means $E_{p_jp_i} \leq T_c - T_p$ and the signal can start to propagate through a latch only after it opens $d_j \geq T_c - T_p$.}
%
This constraint is particularly important when considering an FF with combinational feedback. If no extra latch is added during conversion, the converted circuit would have a single latch $i$ with combinational feedback which violates $C2$. This configuration is dangerous because the transparency phase of the latch must be smaller than the minimum delay of the combinational feedback $\delta_{ii}$ to avoid a hold violation.  
More precisely, the constraint can be formalized as:
\begin{equation*}
\delta_i + \delta_{ii} \geq H_i + T_p.
\end{equation*}
The key point is that this constraint guarantees this configuration is not allowed. In particular, any solution that satisfies this constraint will break such combinational feedback by at least two latches that have non-overlapping clocks. 

A well-known but non-optimal solution to this problem is to convert {\em every} FF into two latches, 
a master and a slave latch, as in \cite{cortadella2006desynchronization,singh2018low}, and retime the slave latches. This master-slave approach satisfies both constraints C1 and C2 but at the cost of doubling the number of sequential elements. That is, before retiming, the extra number of latches added is exactly equal to the number of FFs. 

\subsection{Special Case of Linear Pipelines}

\begin{figure}[tb]
    \centering
    \includegraphics[width=0.4\textwidth]{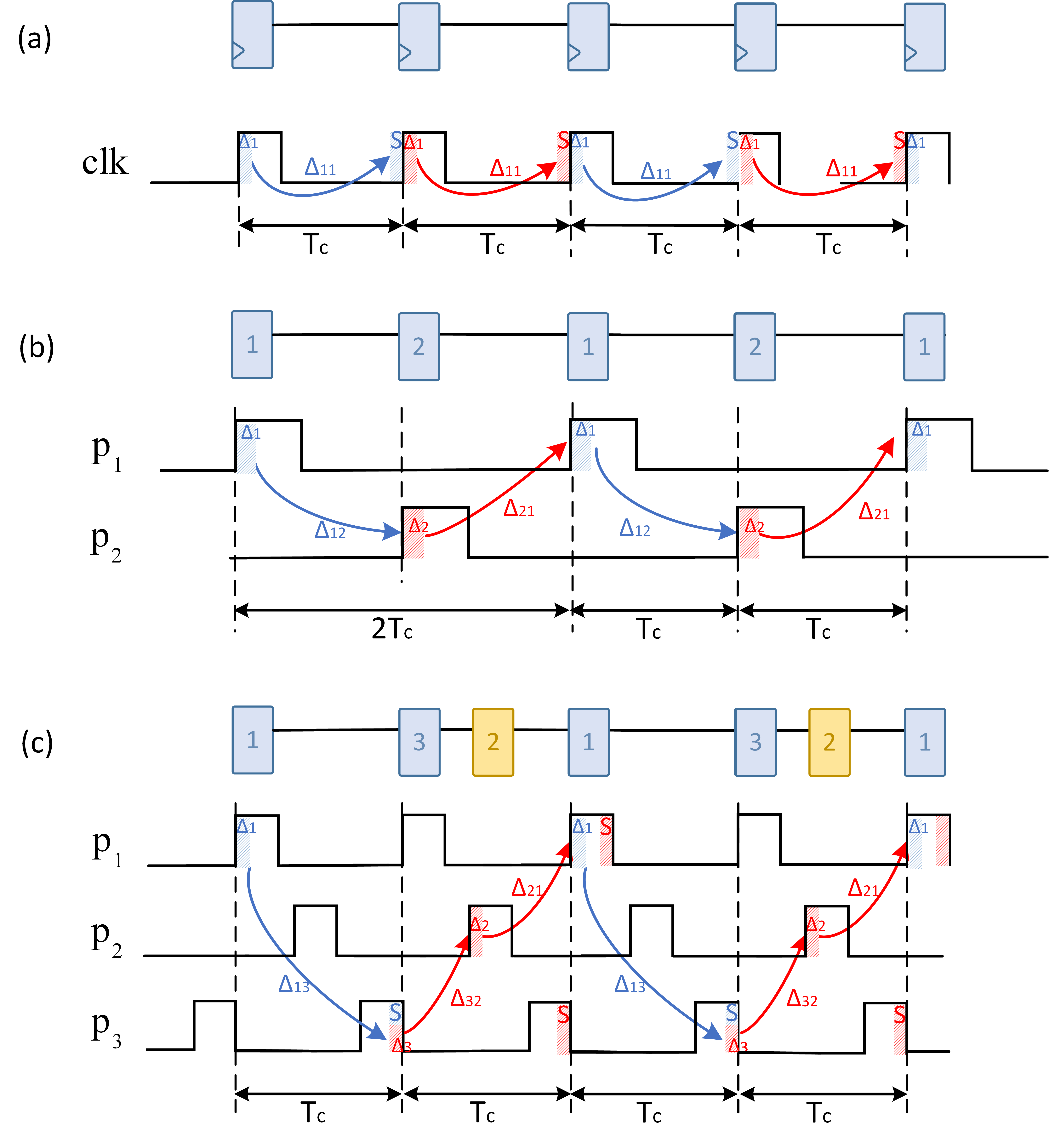}
    \caption{Converting a linear FF-based pipeline (a) to a 2-phase latch-based pipeline (b) and to a 3-phase latch-based pipeline (c)}
    \label{fig:linear_pipeline}
\end{figure}

It is interesting to consider the special case of a linear pipeline because they have no FFs with combinational feedback that must be considered. Such a pipeline is illustrated in Figure \ref{fig:linear_pipeline}(a) and its cycle time $T_c$ is no shorter than $\Delta_{1} + \Delta_{11} + S$, where $\Delta_{1}$ represents the FF's clk-to-q delay, $\Delta_{11}$ represents the longest data-path delay, and $S$ stands for the FF's setup time.

Such linear pipelines can be converted to a latch-based design adding {\em no} extra latches, where we clock alternating pipeline stages with alternating phases of a two-phase non-overlapping clock, as illustrated in Figure \ref{fig:linear_pipeline}(b). 
The problem with this solution is that if each combinational logic stage is critical, the time separation between each phase of the clock must be equal to the original cycle time, i.e., $E_{ij} = T_c$, where $T_c$ represents the original cycle time.  
Letting $T_c^{2P}$ denote the cycle time of the two-phase non-overlapping clocks and assuming $E_{12} = E_{21} = T_c$, Equation \ref{eqn:phase_shift} 
implies 
$e_2 - e_1 = T_c^{2P} + e_1 - e_2 = T_c$ 
and thus, $T_c^{2P} = 2T_c$.
In other words, the frequency of the two-phase clocks must be half that of the original FF-based design.

This analysis highlights the fact that there is a trade-off between the number of extra latches added and the performance of the resulting circuit. To avoid this trivial solution in our formulation, we adopt a third constraint:
\begin{enumerate}[C3:]
    \item the converted latch-based design must have the same throughput as the FF-based design assuming the combinational logic is already critical.
 \end{enumerate}

We can achieve a latch-based design that meets all Constraints C1-C3 in which we add exactly one extra latch stage for every other original pipeline stage using a 3-phase clocked, as illustrated in Figure \ref{fig:linear_pipeline}(c). Notice that as desired, this solution has the same throughput as the original pipeline having phases $p_1$ and $p_3$ open and close their respective latches at the rising edge of the FF-based clock. We rely on the $p_3$ latches time borrowing to properly capture 
near critical combinational paths.
The $p_2$ latches inserted between the $p_3$ and $p_1$ latches prevent data latched by $p_3$ to violate the hold times of the subsequent $p_1$ latches.

\subsection{Optimality}

A natural question to ask is if 3-phase clocking guarantees optimality in terms of the number of required extra latches. This section proves that it is optimal for linear pipelines but does not guarantee optimality for more general non-linear pipelines.


\textbf{Theorem I:} 
At least one latch stage has to be inserted between any 
3 consecutive stages of a linear pipeline.

{\bf Proof by contradiction:} Assume there exists three consecutive stages of a linear pipeline for which no extra latch stage is inserted within the combinational logic between stage 1 and stage 2 or between stage 2 and stage 3.

Let time 0 represent the rising edge of the stage 1 clock. According to Constraints C2 and C3, 
stage 2 clock can only go high during the time window ($T_p$, $T_c - T_p$) and 
must go low no later than $T_c$.

Case 1: Assume stage 1 data is valid at time 0. Since there is no latch between stage 1 and stage 2, stage 2 clock captures data no earlier than $T_c$. Then stage 2 clock should be high during the time period ($T_c - T_p$, $T_c$). 
According to Constraints C2 and C3, stage 3 can only go low during the period ($T_c + T_p$, $2 T_c - T_p$). This means that stage 3 has to capture data before 
time $2 T_c - T_p$. 
Because there is no extra latch inserted between stage 2 and stage 3, stage 3 must capture the data no earlier than $2T_c$. This, however, contradicts the fact that stage 3 must go low before $2T_c - T_p$.

Case 2: Assume the data leaves stage 1 at time $t$ ($0 < t <= T_p$). Then stage 2 needs to sample the data no earlier than time $t + T_c$. This contradicts the fact that stage 2 goes low no later than $T_c$. 
\qedsymbol

\begin{figure}
    \centering
    \includegraphics[width=0.4\textwidth]{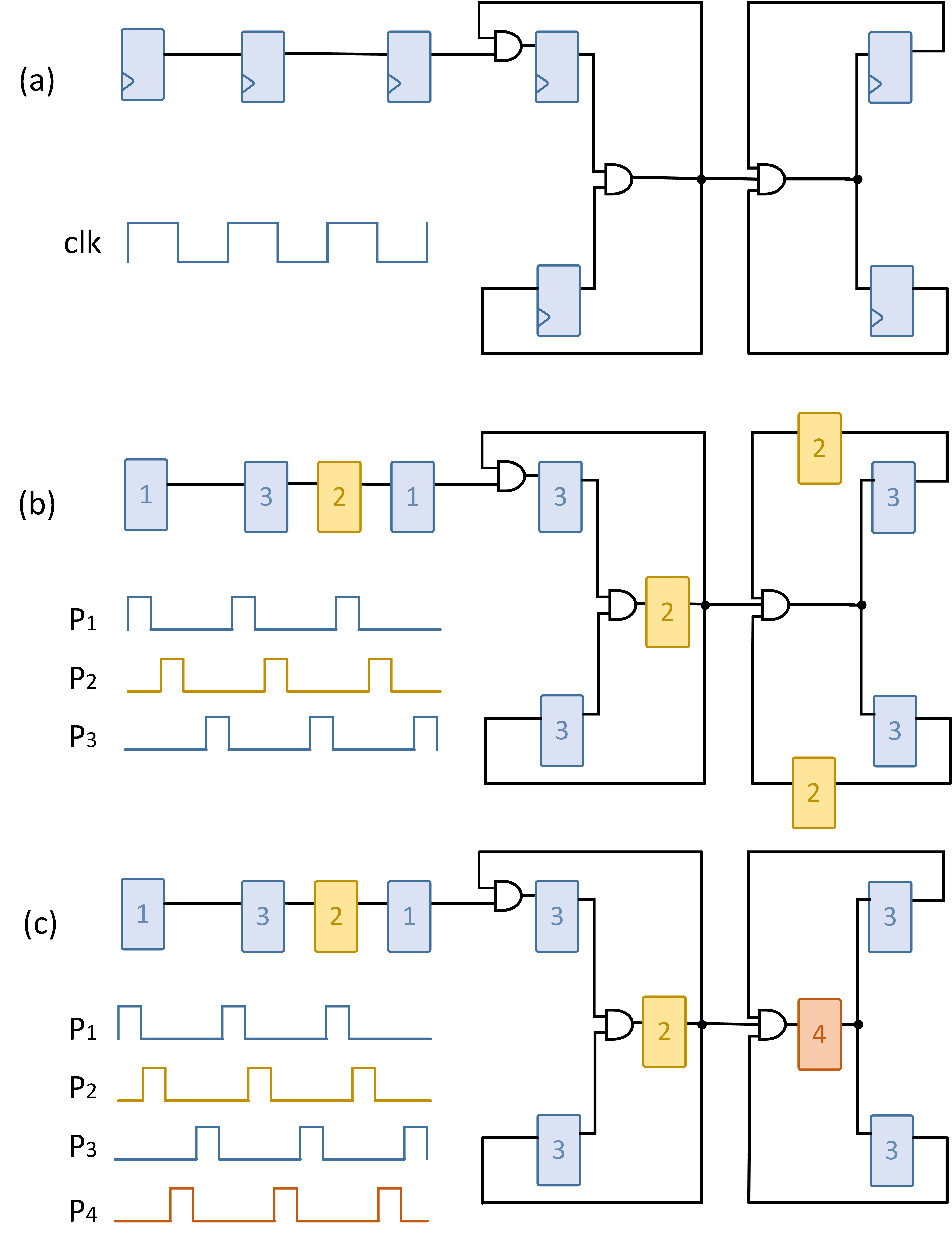}
    \caption{Example non-linear pipeline requiring 4-phase clocking to support the minimum number of extra latches}
    \label{fig:3_4_phase}
\end{figure}

Next, we present Figure \ref{fig:3_4_phase} which illustrates 
an example in which 
4-phase clocking is needed to achieve an optimal latch configuration. 
In particular, Figure \ref{fig:3_4_phase}(a) illustrates 
an original FF-based design where the combinational connections are abstracted to wires for simplicity.
The optimal 3-phase clocking solution requires at least four extra latches,  labeled ``2" in Figure \ref{fig:3_4_phase}(b).
However, 4-phase clocking yields a 
latch-based design that requires only three 
extra latches (labeled 2 and 4 in Figure \ref{fig:3_4_phase}(c)).

To the best of our knowledge, it is an open question as to whether there are optimal latch-based designs that require more than four clock phases. 

Despite this example, the remainder of this paper presents an conversion algorithm that produces three-phase latch-based designs.  
The algorithm is thus not guaranteed to be optimal because it does not support more than three clock phases. It is also not optimal as it considers the restrictive case of adding extra latches only directly after required latches. More specifically, we rely on retiming of these extra latches to position the extra latches within the combinational logic and satisfy constraints C1-C3. 
The separation of these two steps can lead to non-optimal results. Extending our algorithm to support four or more phases and additional latch locations is more complex and is interesting on-going research. 

\section{Conversion algorithm} 
\label{sec:conversion_alg}

Our conversion approach is to automatically decompose the FFs into two groups, ones that will be converted to back-to-back connected latches and ones that will be converted into a single latch. The group of FFs converted to a single latch are assigned to clock phase $p_1$. 
The remaining FFs are converted to latches clocked by either $p_1$ or $p_3$. For this group, an additional latch clocked by $p_2$ is inserted at each latches’ output to create a back-to-back configuration. This means that, by construction, there is no direct data path from $p_3$ to $p_1$ latches.
Min delay related hold problems are avoided by allowing an FF to be assigned to phase $p_1$ and converted to a single latch only if none of its fanout FFs are also assigned to $p_1$. 

\subsection{Integer Linear Programming (ILP)}
\label{sec:ILP}


Each FF is treated as a node $u$ and its $FO(u)$ is the set of FFs that can be reached from the FF $u$ via only combinational logic.
Every node $u$ has two binary parameters, $G(u)$ and $K(u)$.
$G(u)$ decides which group of latches to assign node $u$, either the back-to-back latch group ($G(u) = 1$) or the single-latch group ($G(u)=0$). 
$K(u)$ determines the node $u$'s clock phase, $1$ implies $u$ is clocked by $p_1$ and $0$ implies $u$ is clocked by $p_3$. 
All inserted latches are driven by $p_2$. Our ILP automatically performs this assignment minimizing the number of back-to-back latches as follows:
\begin{equation}
    Minimize \; \sum_{u} G(u)
    \nonumber
\end{equation}
Subject to:
\begin{equation}
\label{eqn:group_assign}
    \forall u \in V: \; \;  G(u) = 
    \begin{cases}
    1, &  K(u) = 0, \\
    1, &  K(u) = 1 \land \exists v \in FO(u) \; K(v) = 1 \\ 
    0, & otherwise 
    \end{cases}
    \nonumber
\end{equation}
\begin{equation}
\label{eqn:k_assign}
    K(u) = 
    \begin{cases}
    1, &  \forall u \in PI \\ 
    \{0, 1\}, &  \forall u \in V
    \end{cases}
    \nonumber
\end{equation}

Here $PI$ stands for the set of all primary input ports and set $V$ contains all nodes in the circuit. To provide consistency to the interface of the design, we assign all primary input ports (\textit{PI}s) as if they were clocked by $p_1$.
 
To make the ILP compatible with Gurobi  \cite{gurobi}, we convert the conditional equations into inequalities:

\begin{equation*}
    Minimize \; \sum_{u} G(u)
\end{equation*}
Subject to:
\begin{equation}
\begin{cases}
    G(u) + K(u) \geq 1  &  \forall u \in V\\
    G(u) \geq K(u) + K(v) - 1 & \forall u \in V, \forall v \in FO(u) \\
    G(u) \geq K(v) & \forall u \in PI, \forall v \in FO(u) 
\end{cases}
\nonumber
\end{equation}
\begin{equation*}
    G(u), K(u) \in \{0,1\}
\end{equation*}
The first constraint that implies when $K(u)=0$ inequality $G(u) \geq 1$ is satisfied is corresponding to the first condition in \eqref{eqn:group_assign}.
The second constraint makes sure $G(u)=1$ if $K(u)$ and any of its fanout $K(v)$ are both 1, rephrasing the second condition in \eqref{eqn:group_assign}.
Applying the assumption that all PIs are clocked by $p_1$ to the second constraint above, we obtain the third inequality.

\subsection{The Design Flow}
\label{sec:design_flow}

The ILP described in the last section is the core step in a design flow that supports FF-based to 3-phase latch-based design conversion.
The first step of our design flow is to run standard synchronous synthesis on the given FF-based RTL design. 
Here, we take care to enable clock gating to minimize the number of FFs with self-loops which would otherwise unduly constrain the optimization problem.
\begin{figure}
    \centering
    \includegraphics[width=0.45\textwidth]{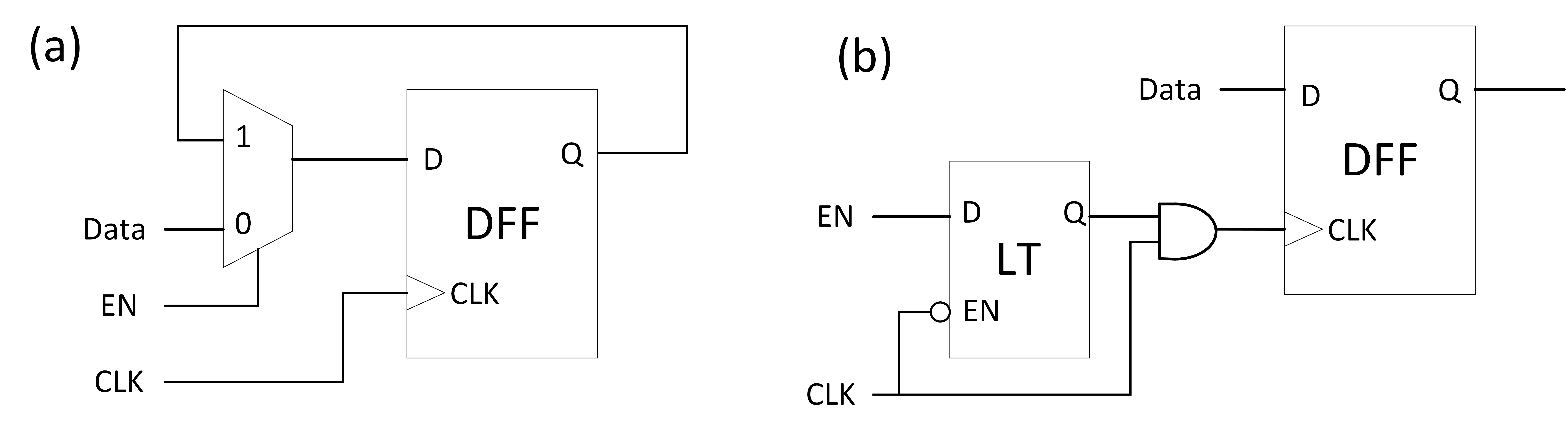}
    \caption{Enabled (a) to gated clock (b) transformation}
    \label{fig:CLK_GATE}
\end{figure}
To be specific, the gated clock, shown in Figure \ref{fig:CLK_GATE}(b), is set to be the preferred clock gating style, as compared to enabled clocks illustrated in Figure \ref{fig:CLK_GATE}(a).

Using Python and TCL scripts that interface a leading commercial logic synthesis tool to the Gurobi Integer Linear Program solver \cite{gurobi}, we then take the resulting FF-based design, identify the connections between FFs, and formulate the ILP described in Section \ref{sec:ILP}. We run the ILP, and, using the results, create the equivalent 3-phase latch-based synchronous design by defining the three-phase clocks and connecting them to their associated latches. 

%
\begin{figure}
    \centering
    \includegraphics[width=0.45\textwidth]{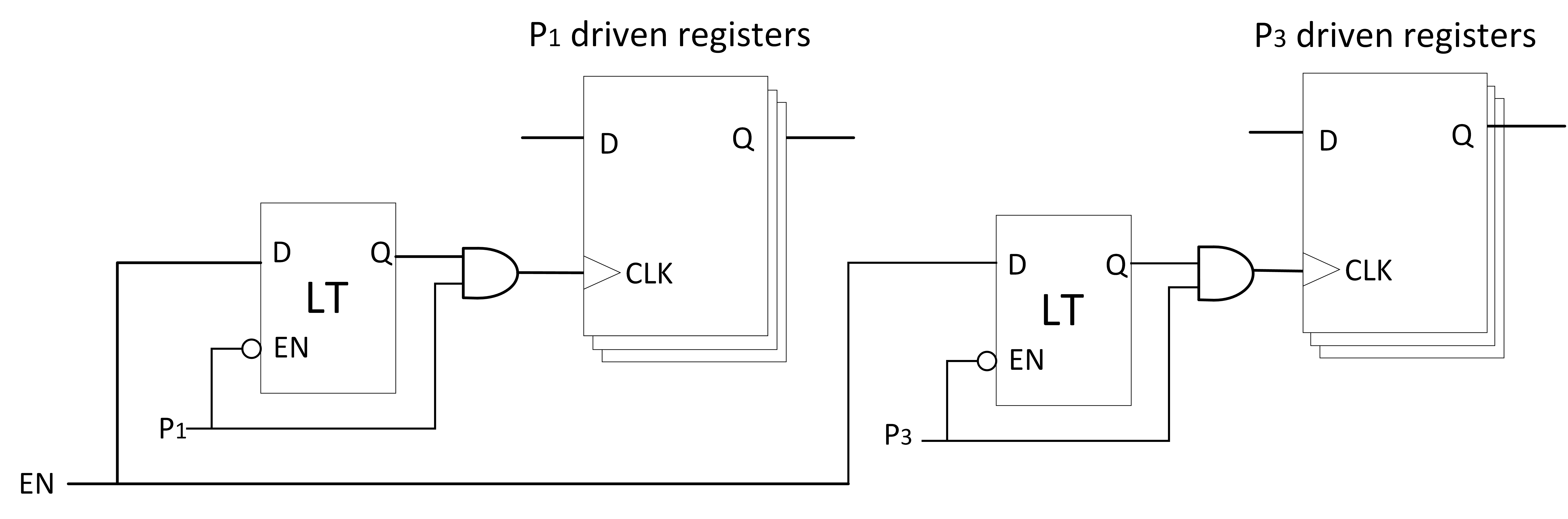}
    \caption{Duplicated clock gating logic for phase conversion}
    \label{fig:CLK_GATE_3phase}
\end{figure}
For each latch that are clock gated, we trace the clock signal back through the clock gating logic and replace the clock with $p1$ or $p3$. 
In the case of latches belonging to the same clock gating register bank but driven by different clock phases, the clock gating logic is duplicated and connected to the two clock phases separately, as shown in Figure \ref{fig:CLK_GATE_3phase}. 
We then retime the newly added latches, as described below. 
The last step in the design flow, left as future work in this paper, 
is the physical design step which includes implementation of the three-phase clock trees.

\subsection{Modified Retiming}
\label{sec:retiming}

Retiming re-positions the added latches within the combinational logic minimizing area while satisfying all latch constraints.
%
Unfortunately, many commercial tools have limited support for retiming latches. They do, however, have well-optimized support for the retiming of FFs.
Using this fact, \cite{chinnery2002closing} proposed to retime latches by mapping it to an FF-based retiming problem.  
Given a synthesized design with clock period $T_c$, they replace each FF with two FFs and retime the 
entire design with a faster clock constraint of 
half the original period ($T_c/2$). 
After splitting the combinational logic, the 
FFs are converted into alternating
transparent low and high latches.

\begin{figure}[t]
    \centering
    \includegraphics[width=0.25\textwidth]{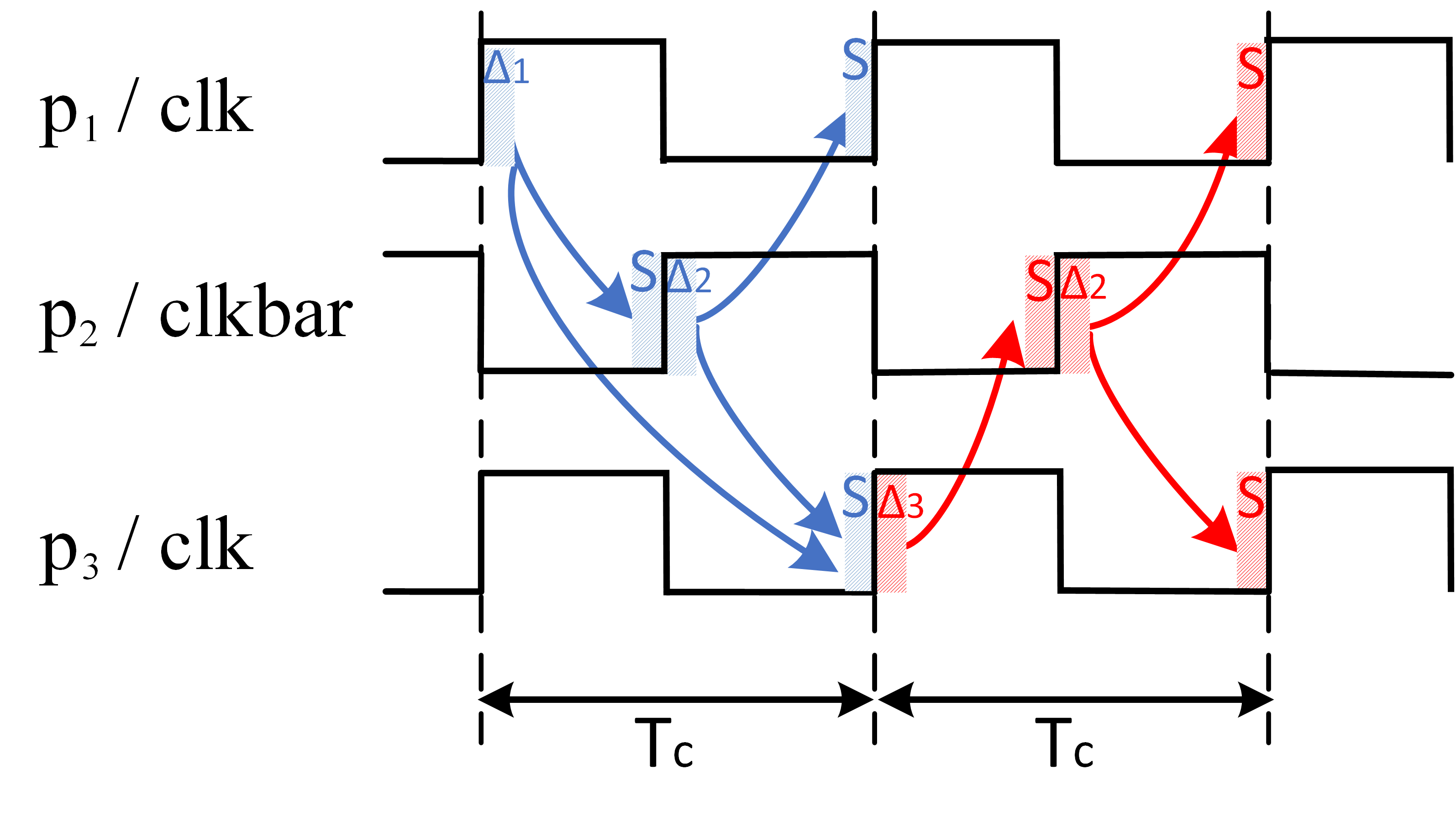}
    \caption{3-phase clocks for modified retiming}
    \label{fig:retiming_3phase}
\end{figure}

In this paper, instead of halving the cycle, we keep the cycle time unchanged but use back-to-back FFs, where the first FF is controlled by clk and the second clocked by clk inverted (clkbar).
The group that is converted to a single latch is replaced with a single FF, also controlled by clk.
The 3-phase clocks are mapped to clk and clkbar as shown in Figure \ref{fig:retiming_3phase}. 
Phase $p_1$ and $p_3$ are mapped to clk and $p_2$ is tied to clkbar.
We then retime the circuit only allowing FFs tied to clkbar to move.
This splits the combinational logic in the pipeline stages that require an extra latch into two with each part being able to operate at twice the frequency (cycle time $T_c/2$). 

After the relocation of FFs clocked by clkbar, all FFs can be converted back to latches with their designated 3-phase assignments. Further optimization is then triggered to optimize the sizes of gates in the retimed latch-based design.

\section{Experimental Results}
\label{sec:results}

This section quantifies the benefits of the proposed conversion algorithm comparing the resulting 3-phase design to the original FF-based as well as traditional master-slave latch-based designs. 
The experiments rely on an industrial 28-nm FDSOI CMOS cell library and a range of circuits that include, 
ISCAS89 benchmark circuits \cite{ISCAS}, CEP submodules \cite{CEP}, and three CPU designs, a 3-stage MIPS Open Core Plasma \cite{plasma}, a RISC-V Rocket Core \cite{rocket}, and an ARM-M0 core \cite{armm0}.
We validated both master-slave and 3-phase latch-based circuits by streaming inputs to the FF-based and latch-based designs and compare output streams.\footnote{For ISCAS designs we used auto-generated pseudo-random input streams. For CEP and CPU designs, we used the open-source provided testbenches. In particular, Plasma was running the ``pi" program, ARM-M0 was running the ``hello world" program, RISC-V was running the ``rv32ui-v-simple" program, and CEP designs were running the open-source provided self-check programs.}
These gate-level simulations were also used to determine signal activity used to measure the relative power consumption of our approach. 
Note, however, that because our results are post-synthesis, our analysis does not consider 
the power consumption of the clock trees.
All experiments were run on two Intel Xeon E5-2450 v2 CPUs with 128GB of RAM.

Note that for a fair comparison, all designs are run at the same frequency and the modified work-around retiming strategy described in Section \ref{sec:conversion_alg} is also performed on the master-slave latch-based designs. 

Table \ref{tab:num_reg_results} summarizes the number of registers (FFs/latches) in the original FF-based, conventional master-slave latch-based, and 3-phase latch-based designs. 
The right most two columns show the savings of our approach in terms of the number of latches in 3-phase latch-based designs compared to the doubled number of FFs in FF-based and the number of latches in master-slave latch-based designs, respectively.

\begin{table}[t]
\scriptsize\addtolength{\tabcolsep}{0pt}
    \centering
    \begin{tabular}{|c|c|c|c|c|c|c|}
    \hline
     \multicolumn{2}{|c|}{\multirow{2}{*}{Design}} & \multirow{2}{*}{FF} & \multirow{2}{*}{M-S} & \multirow{2}{*}{3-phase} & \multicolumn{2}{c|}{Save (\%)} \\ \cline{6-7} 
     \multicolumn{2}{|c|}{} & & & & 2*FF & M-S\\ \hline
\multirow{12}{*}{ISCAS}												
&s1196	&	18	&	36	&	26	&	27.8	&	27.8	\\	\cline{2-7}
&s1238	&	18	&	36	&	26	&	27.8	&	27.8	\\	\cline{2-7}
&s1423	&	74	&	158	&	167	&	-12.8	&	-5.7	\\	\cline{2-7}
&s1488	&	6	&	12	&	12	&	0.0	&	0.0	\\	\cline{2-7}
&s5378	&	164	&	326	&	250	&	23.8	&	23.3	\\	\cline{2-7}
&s9234	&	145	&	299	&	257	&	11.4	&	14.0	\\	\cline{2-7}
&s13207	&	460	&	905	&	761	&	17.3	&	15.9	\\	\cline{2-7}
&s15850	&	449	&	922	&	818	&	8.9	&	11.3	\\\cline{2-7}	
&s35932	&	1728	&	3456	&	2738	&	20.8	&	20.8	\\	\cline{2-7}
&s38417	&	1490	&	2953	&	2466	&	17.2	&	16.5	\\	\cline{2-7}
&s38584	&	1268	&	2621	&	2478	&	2.3	&	5.5	\\	\cline{2-7}
\rowcolor[gray]{.9}\cellcolor{white}												
&Average	&	529.1	&	1065.8	&	909.0	&	14.1	&	14.7	\\	\hline
\multirow{5}{*}{CEP}												
&AES	&	9703	&	17760	&	13578	&	30.0	&	23.5	\\	\cline{2-7}
&DES3	&	425	&	861	&	594	&	30.1	&	31.0	\\	\cline{2-7}
&SHA256	&	1554	&	3133	&	2581	&	17.0	&	17.6	\\	\cline{2-7}
&MD5	&	782	&	1586	&	1086	&	30.6	&	31.5	\\	\cline{2-7}
\rowcolor[gray]{.9}\cellcolor{white}												
&Average&		3116.0	&	5835.0	&	4459.8	&	28.4	&	23.6	\\	\hline
\multirow{4}{*}{CPU}												
&Plasma	&	1554	&	3159	&	2150	&	30.8	&	31.9	\\	\cline{2-7}
&RISC-V	&	2561	&	5226	&	4178	&	18.4	&	20.1	\\	\cline{2-7}
&ARM-M0	&	1334	&	2738	&	2185	&	18.1	&	20.2	\\	\cline{2-7}
\rowcolor[gray]{.9}\cellcolor{white}												
&Average&		1816.3	&	3707.7	&	2837.7	&	21.9	&	23.5	\\	\hline
\rowcolor[gray]{.9}												
\multicolumn{2}{|c|}{Average}&		1318.5	&	2565.9	&	2019.5	&	23.4	&	21.3	\\	\hline	
    \end{tabular}
    \caption{Number of registers (FFs or latches) in the original flip-flop (FF), converted master-slave latch (M-S), and proposed 3-phase latch based designs}
    \label{tab:num_reg_results}
\end{table}

\begin{table*}[bp]
\scriptsize\addtolength{\tabcolsep}{-3pt}
    \centering
    \begin{tabular}{|c|c|c|c|c|c|c|c|c|c|c|c|c|c|c|c|c|}
    \hline
     \multicolumn{2}{|c|}{\multirow{2}{*}{Design}} & \multicolumn{3}{c|}{FF area} & \multicolumn{3}{c|}{M-S area} & \multicolumn{3}{c|}{3-phase area} & \multicolumn{3}{c|}{Save (\%) wrt. FF} & \multicolumn{3}{c|}{Save (\%) wrt. M-S}\\ \cline{3-17}
     \multicolumn{2}{|c|}{} & Comb & Seq & Total & Comb & Seq & Total & Comb & Seq & Total & Comb & Seq & Total & Comb & Seq & Total \\ \hline
\multirow{12}{*}{ISCAS}																																
&s1196	&	172.7	&	67.6	&	240.2	&	163.4	&	58.8	&	222.1	&	167.9	&	44.1	&	212.0	&	2.7	&	34.8	&	11.8	&	-2.8	&	25.0	&	4.6	\\	\cline{2-17}
&s1238	&	168.6	&	67.6	&	236.2	&	161.1	&	58.8	&	219.8	&	162.1	&	44.1	&	206.1	&	3.9	&	34.8	&	12.7	&	-0.6	&	25.0	&	6.2	\\	\cline{2-17}
&s1423	&	210.5	&	292.8	&	503.3	&	212.2	&	269.6	&	481.8	&	205.1	&	270.4	&	475.6	&	2.6	&	7.6	&	5.5	&	3.3	&	-0.3	&	1.3	\\	\cline{2-17}
&s1488	&	194.5	&	22.5	&	217.1	&	188.2	&	19.9	&	208.1	&	191.9	&	19.6	&	211.5	&	1.3	&	13.0	&	2.6	&	-2.0	&	1.6	&	-1.6	\\	\cline{2-17}
&s5378	&	396.3	&	615.6	&	1011.8	&	387.3	&	532.0	&	919.3	&	388.1	&	420.7	&	808.8	&	2.1	&	31.7	&	20.1	&	-0.2	&	20.9	&	12.0	\\	\cline{2-17}
&s9234	&	287.7	&	557.5	&	845.2	&	293.1	&	494.3	&	787.4	&	271.2	&	423.2	&	694.4	&	5.7	&	24.1	&	17.8	&	7.5	&	14.4	&	11.8	\\	\cline{2-17}
&s13207	&	551.5	&	1750.2	&	2301.6	&	531.9	&	1486.4	&	2018.3	&	584.4	&	1272.1	&	1856.6	&	-6.0	&	27.3	&	19.3	&	-9.9	&	14.4	&	8.0	\\	\cline{2-17}
&s15850	&	842.0	&	1723.9	&	2565.8	&	923.4	&	1513.7	&	2437.1	&	815.0	&	1353.1	&	2168.1	&	3.2	&	21.5	&	15.5	&	11.7	&	10.6	&	11.0	\\\cline{2-17}	
&s35932	&	3087.9	&	6486.2	&	9574.1	&	3046.1	&	5640.2	&	8686.3	&	3056.2	&	4585.6	&	7641.8	&	1.0	&	29.3	&	20.2	&	-0.3	&	18.7	&	12.0	\\	\cline{2-17}
&s38417	&	2622.6	&	5650.8	&	8273.4	&	2787.8	&	4835.1	&	7622.9	&	3042.2	&	4110.7	&	7152.9	&	-16.0	&	27.3	&	13.5	&	-9.1	&	15.0	&	6.2	\\	\cline{2-17}
&s38584	&	3169.2	&	4905.0	&	8074.2	&	3225.6	&	4307.3	&	7533.0	&	3227.6	&	4051.0	&	7278.6	&	-1.8	&	17.4	&	9.9	&	-0.1	&	6.0	&	3.4	\\	\cline{2-17}
\rowcolor[gray]{.9}\cellcolor{white}																																
&Average	&	1063.9	&	2012.7	&	3076.6	&	1083.6	&	1746.9	&	2830.6	&	1101.1	&	1508.6	&	2609.7	&	-3.5	&	25.0	&	15.2	&	-1.6	&	13.6	&	7.8	\\	\hline
\multirow{5}{*}{CEP}																																
&AES	&	102418.8	&	25367.3	&	127786.1	&	94989.6	&	26086.9	&	121076.4	&	97769.5	&	19943.4	&	117712.9	&	4.5	&	21.4	&	7.9	&	-2.9	&	23.6	&	2.8	\\	\cline{2-17}
&DES3	&	1462.8	&	1127.4	&	2590.2	&	1503.7	&	1266.6	&	2770.3	&	1467.5	&	872.5	&	2340.0	&	-0.3	&	22.6	&	9.7	&	2.4	&	31.1	&	15.5	\\	\cline{2-17}
&SHA256	&	4383.4	&	4131.2	&	8514.6	&	4495.8	&	4630.3	&	9126.1	&	4285.0	&	3791.0	&	8076.0	&	2.2	&	8.2	&	5.2	&	4.7	&	18.1	&	11.5	\\	\cline{2-17}
&MD5	&	4270.8	&	2117.9	&	6378.0	&	3907.0	&	2406.9	&	6313.9	&	3837.3	&	1612.4	&	5449.7	&	10.1	&	23.9	&	14.6	&	1.8	&	33.0	&	13.7	\\	\cline{2-17}
\rowcolor[gray]{.9}\cellcolor{white}																																
&Average	&	28133.9	&	8186.0	&	36317.2	&	26224.0	&	8597.7	&	34821.7	&	26839.8	&	6554.8	&	33394.6	&	4.6	&	19.9	&	8.0	&	-2.3	&	23.8	&	4.1	\\	\hline
\multirow{4}{*}{CPU}																																
&Plasma	&	3998.6	&	4749.9	&	8748.5	&	4192.3	&	4879.2	&	9071.5	&	4781.4	&	3356.4	&	8137.8	&	-19.6	&	29.3	&	7.0	&	-14.1	&	31.2	&	10.3	\\	\cline{2-17}
&RISC-V	&	7115.0	&	6860.6	&	13975.6	&	7710.2	&	7765.1	&	15475.3	&	7929.1	&	6138.3	&	14067.4	&	-11.4	&	10.5	&	-0.7	&	-2.8	&	20.9	&	9.1	\\	\cline{2-17}
&ARM-M0	&	6458.3	&	3985.2	&	10443.5	&	6821.9	&	4639.4	&	11461.4	&	6655.1	&	3326.2	&	9981.3	&	-3.0	&	16.5	&	4.4	&	2.4	&	28.3	&	12.9	\\	\cline{2-17}
\rowcolor[gray]{.9}\cellcolor{white}																																
&Average	&	5857.3	&	5198.6	&	11055.9	&	6241.5	&	5761.2	&	12002.7	&	6455.2	&	4273.6	&	10728.8	&	-10.2	&	17.8	&	3.0	&	-3.4	&	25.8	&	10.6	\\	\hline
\rowcolor[gray]{.9}																																
\multicolumn{2}{|c|}{Average}	&	7878.4	&	3915.5	&	11793.3	&	7530.0	&	3938.4	&	11468.4	&	7713.2	&	3090.8	&	10804.0	&	2.1	&	21.1	&	8.4	&	-2.4	&	21.5	&	5.8	\\	\hline
    \end{tabular}
    \caption{Areas ($\mu m^2$) of flip-flop (FF), master-slave latch (M-S), and 3-phase latch-based designs}
    \label{tab:area_results}
\end{table*}

The results show that the proposed algorithm reduces the number of latches by an average of 23.4\% and 21.3\% compared to FF-based and master-slave latch-based designs, respectively.
Notice that the 3-phase algorithm has the least overall benefit on the ISCAS89 circuits and, in particular, no benefit on s1488 and s1423.
According to \cite{brglez1989combinational}, s1488 is re-synthesized from a controller and may suggest that our algorithm brings limited benefits to control dominated designs that have a predominance of FFs with combinational feedback.

\begin{table*}[htbp]
\scriptsize\addtolength{\tabcolsep}{0pt}
    \centering
    \begin{tabular}{|c|c|c|c|c|c|c|c|c|c|c|c|c|}
    \hline
     \multicolumn{2}{|c|}{\multirow{2}{*}{Design}} & \multicolumn{3}{c|}{FF power} & \multicolumn{3}{c|}{M-S power} & \multicolumn{3}{c|}{3-phase power} & \multicolumn{2}{c|}{Total Save (\%) }\\ \cline{3-13}
     \multicolumn{2}{|c|}{} & Comb & Seq & Total & Comb & Seq & Total & Comb & Seq & Total & FF vs 3-P & M-S vs 3-P \\ \hline
\multirow{12}{*}{ISCAS}																								
&s1196	&	0.20	&	0.12	&	0.32	&	0.19	&	0.07	&	0.26	&	0.21	&	0.05	&	0.26	&	20.36	&	1.60	\\	\cline{2-13}
&s1238	&	0.20	&	0.12	&	0.32	&	0.20	&	0.07	&	0.27	&	0.21	&	0.05	&	0.26	&	19.07	&	3.83	\\	\cline{2-13}
&s1423	&	0.29	&	0.38	&	0.66	&	0.15	&	0.25	&	0.43	&	0.15	&	0.26	&	0.41	&	37.39	&	4.49	\\	\cline{2-13}
&s1488	&	0.18	&	0.04	&	0.22	&	0.20	&	0.02	&	0.23	&	0.20	&	0.02	&	0.22	&	0.99	&	2.47	\\	\cline{2-13}
&s5378	&	0.37	&	1.00	&	1.37	&	0.35	&	0.56	&	0.91	&	0.40	&	0.48	&	0.88	&	35.61	&	3.78	\\	\cline{2-13}
&s9234	&	0.17	&	0.62	&	0.79	&	0.08	&	0.43	&	0.53	&	0.08	&	0.38	&	0.46	&	42.28	&	14.84	\\	\cline{2-13}
&s13207	&	0.44	&	2.19	&	2.63	&	0.32	&	1.44	&	1.78	&	0.33	&	1.28	&	1.61	&	38.83	&	9.46	\\	\cline{2-13}
&s15850	&	2.28	&	0.44	&	2.72	&	0.56	&	1.25	&	1.88	&	0.44	&	1.39	&	1.83	&	32.91	&	3.09	\\	\cline{2-13}
&s35932	&	12.87	&	2.74	&	15.62	&	2.70	&	8.59	&	11.30	&	2.89	&	7.48	&	10.36	&	33.64	&	8.26	\\	\cline{2-13}
&s38417	&	6.08	&	2.05	&	8.13	&	2.22	&	3.29	&	5.55	&	2.36	&	2.68	&	5.04	&	37.92	&	9.15	\\	\cline{2-13}
&s38584	&	9.56	&	3.31	&	12.87	&	4.63	&	6.23	&	11.14	&	3.67	&	5.00	&	8.67	&	32.63	&	22.18	\\	\cline{2-13}
\rowcolor[gray]{.9}\cellcolor{white}																								
&Average	&	2.97	&	1.18	&	4.15	&	1.06	&	2.02	&	3.12	&	0.99	&	1.73	&	2.73	&	34.28	&	12.52	\\	\hline
	\multirow{5}{*}{CEP}																							
&AES	&	0.22	&	12.64	&	12.86	&	0.19	&	4.02	&	4.21	&	0.17	&	3.02	&	3.19	&	75.21	&	24.35	\\	\cline{2-13}
&DES3	&	0.48	&	0.31	&	0.79	&	0.47	&	0.27	&	0.74	&	0.40	&	0.16	&	0.56	&	28.41	&	24.04	\\	\cline{2-13}
&SHA256	&	0.17	&	0.11	&	0.27	&	0.12	&	0.39	&	0.52	&	0.12	&	0.25	&	0.38	&	-37.78	&	27.39	\\	\cline{2-13}
&MD5	&	0.29	&	0.05	&	0.34	&	0.27	&	0.19	&	0.47	&	0.16	&	0.12	&	0.28	&	16.50	&	40.15	\\	\cline{2-13}
\rowcolor[gray]{.9}\cellcolor{white}																								
&Average	&	0.29	&	3.27	&	3.56	&	0.26	&	1.22	&	1.49	&	0.21	&	0.89	&	1.10	&	69.08	&	25.82	\\	\hline
	\multirow{4}{*}{CPU}																							
&Plasma	&	0.84	&	0.81	&	1.65	&	1.06	&	0.72	&	1.81	&	0.64	&	0.52	&	1.16	&	29.75	&	35.97	\\	\cline{2-13}
&RISC-V	&	0.54	&	0.32	&	0.86	&	1.03	&	0.42	&	1.48	&	0.36	&	0.66	&	1.02	&	-18.01	&	31.06	\\	\cline{2-13}
&ARM-M0	&	1.25	&	0.76	&	2.01	&	0.69	&	1.34	&	2.05	&	1.05	&	0.51	&	1.56	&	22.73	&	23.90	\\	\cline{2-13}
\rowcolor[gray]{.9}\cellcolor{white}																								
&Average	&	0.88	&	0.63	&	1.51	&	0.93	&	0.83	&	1.78	&	0.68	&	0.56	&	1.25	&	17.54	&	29.98	\\	\hline
\rowcolor[gray]{.9}																								
\multicolumn{2}{|c|}{Average}	&	2.02	&	1.56	&	3.58	&	0.86	&	1.64	&	2.53	&	0.77	&	1.35	&	2.12	&	40.80	&	16.30	\\	\hline	
    \end{tabular}
    \caption{Power consumption (mW) based on simulation in the original flip-flop (FF), converted master-slave latch (M-S), and proposed 3-phase (3-P) latch-based designs}
    \label{tab:power_results}
\end{table*}

Table \ref{tab:area_results} shows the areas of combinational, sequential logic, and the total for each benchmark for FF, master-slave, and 3-phase designs. It also shows the percentage area reductions for the 3-phase
designs when compared to both the FF- and master-slave designs.
According to the table, the 3-phase designs achieve an average of 8.4\% and 5.8\% savings in total area compared to FF-based and master-slave latch-based designs, respectively.
Notice that the three CPU benchmarks show a relatively high area reduction over master-slave designs but a relatively low area saving compared to FF-based designs. This is a result of the fact that converting FF- to latch-based designs sometimes increases the combinational logic area depending on the results of retiming. In particular, for the CPU designs, the average area of combinational logic increases by 10.2\% and 3.4\% for 3-phase compared to FF-based and master-slave latch-based designs. 
On the other hand, the area of the combinational logic changes less in the ISCAS and CEP designs. To be specific, the combinational logic area of ISCAS and CEP 3-phase designs are increased by 3.5\% and decreased by 4.6\% with respect to FF-based designs and increased by an average of 1.6\% and 2.3\% over master-slave latch-based designs, respectively.
Note the degree of logic area increase is clock-frequency dependent and re-running these experiments at lower frequencies, reduces this impact. 

Table \ref{tab:power_results} reports the power dissipation of the resulting designs based on the specific signal activities determined by our back-annotated gate-level simulations.
The 3-phase latch-based designs show an average power reduction of 40.8\% compared to the FF-based designs and 16.3\% compared to the master-slave latch-based designs. 
The table shows that the proposed approach can save up to 75\% of the power consumption at the same frequency when compared to traditional FF-based designs. 
\begin{table*}[htbp]
\scriptsize\addtolength{\tabcolsep}{0pt}
    \centering
    \begin{tabular}{|c|c|c|c|c|c|c|c|c|c|c|c|c|}
    \hline
     \multicolumn{2}{|c|}{\multirow{2}{*}{Design}} & \multicolumn{3}{c|}{FF power} & \multicolumn{3}{c|}{M-S power} & \multicolumn{3}{c|}{3-phase power} & \multicolumn{2}{c|}{Total Save (\%) }\\ \cline{3-13}
     \multicolumn{2}{|c|}{} & Comb & Seq & Total & Comb & Seq & Total & Comb & Seq & Total & FF vs 3-P & M-S vs 3-P \\ \hline
\multirow{12}{*}{ISCAS}																								
&s1196	&	0.12	&	0.09	&	0.21	&	0.12	&	0.05	&	0.16	&	0.11	&	0.03	&	0.15	&	29.53	&	10.01	\\	\cline{2-13}
&s1238	&	0.11	&	0.09	&	0.21	&	0.12	&	0.05	&	0.16	&	0.11	&	0.04	&	0.15	&	28.73	&	10.77	\\	\cline{2-13}
&s1423	&	0.26	&	0.34	&	0.60	&	0.16	&	0.22	&	0.37	&	0.13	&	0.21	&	0.35	&	42.08	&	7.61	\\	\cline{2-13}
&s1488	&	0.16	&	0.04	&	0.20	&	0.17	&	0.02	&	0.19	&	0.16	&	0.02	&	0.18	&	8.26	&	3.66	\\	\cline{2-13}
&s5378	&	0.48	&	0.83	&	1.32	&	0.46	&	0.44	&	0.89	&	0.47	&	0.37	&	0.84	&	36.47	&	6.30	\\	\cline{2-13}
&s9234	&	0.26	&	0.62	&	0.88	&	0.19	&	0.42	&	0.61	&	0.17	&	0.37	&	0.53	&	39.54	&	12.67	\\	\cline{2-13}
&s13207	&	0.53	&	2.11	&	2.64	&	0.36	&	1.35	&	1.71	&	0.38	&	1.17	&	1.54	&	41.47	&	9.75	\\	\cline{2-13}
&s15850	&	0.85	&	1.86	&	2.71	&	0.62	&	1.37	&	1.99	&	0.54	&	1.20	&	1.74	&	35.66	&	12.37	\\	\cline{2-13}
&s35932	&	2.99	&	9.54	&	12.53	&	2.95	&	5.86	&	8.81	&	2.79	&	4.73	&	7.51	&	40.04	&	14.71	\\	\cline{2-13}
&s38417	&	2.64	&	4.82	&	7.46	&	3.03	&	4.56	&	7.59	&	2.25	&	3.18	&	5.43	&	27.17	&	28.44	\\	\cline{2-13}
&s38584	&	3.38	&	5.04	&	8.42	&	3.07	&	4.55	&	7.62	&	2.22	&	3.52	&	5.74	&	31.91	&	24.76	\\	\cline{2-13}
\rowcolor[gray]{.9}\cellcolor{white}																								
&Average	&	1.07	&	2.31	&	3.38	&	1.02	&	1.72	&	2.74	&	0.85	&	1.35	&	2.20	&	35.00	&	19.78	\\	\hline
	\multirow{5}{*}{CEP}																							
&AES	&	24.50	&	18.69	&	43.19	&	24.61	&	11.41	&	36.02	&	21.55	&	7.12	&	28.67	&	33.62	&	20.42	\\	\cline{2-13}
&DES3	&	0.78	&	0.74	&	1.53	&	0.68	&	0.40	&	1.08	&	0.64	&	0.24	&	0.89	&	41.84	&	17.44	\\	\cline{2-13}
&SHA256	&	2.01	&	2.82	&	4.83	&	1.54	&	1.70	&	3.24	&	1.42	&	1.18	&	2.60	&	46.07	&	19.69	\\	\cline{2-13}
&MD5	&	2.26	&	1.33	&	3.59	&	1.97	&	0.88	&	2.85	&	1.33	&	0.49	&	1.82	&	49.30	&	36.18	\\	\cline{2-13}
\rowcolor[gray]{.9}\cellcolor{white}																								
&Average	&	7.39	&	5.90	&	13.28	&	7.20	&	3.60	&	10.80	&	6.24	&	2.26	&	8.49	&	36.04	&	21.33	\\	\hline
	\multirow{4}{*}{CPU}																							
&Plasma	&	1.57	&	1.67	&	3.24	&	1.92	&	2.63	&	4.55	&	1.67	&	1.45	&	3.12	&	3.50	&	31.44	\\	\cline{2-13}
&RISC-V	&	2.60	&	2.48	&	5.08	&	2.80	&	3.26	&	6.06	&	2.21	&	2.10	&	4.31	&	15.23	&	28.85	\\	\cline{2-13}
&ARM-M0	&	1.75	&	1.03	&	2.79	&	2.25	&	1.72	&	3.97	&	1.93	&	0.96	&	2.89	&	-3.66	&	27.30	\\	\cline{2-13}
\rowcolor[gray]{.9}\cellcolor{white}																								
&Average	&	1.98	&	1.73	&	3.70	&	2.32	&	2.54	&	4.86	&	1.94	&	1.50	&	3.44	&	7.07	&	29.24	\\	\hline
\rowcolor[gray]{.9}																								
\multicolumn{2}{|c|}{Average}	&	2.63	&	3.01	&	5.63	&	2.61	&	2.27	&	4.88	&	2.23	&	1.58	&	3.80	&	32.49	&	22.11	\\	\hline
    \end{tabular}
    \caption{Power consumption (mW) based on switching activity in the original flip-flop (FF), converted master-slave latch (M-S), and proposed 3-phase (3-P) latch-based designs}
    \label{tab:activity_power_results}
\end{table*}
\begin{table}[tb!]
\scriptsize\addtolength{\tabcolsep}{-2pt}
    \centering
    \begin{tabular}{|c|c|c|c|c|c|c|}
    \hline
     \multicolumn{2}{|c|}{\multirow{2}{*}{Design}} & \multirow{2}{*}{FF Total} & 
     \multirow{2}{*}{M-S Total} & 
     \multicolumn{3}{c|}{3-Phase} \\
     \cline{5-7} 
 \multicolumn{2}{|c|}{} & & & ILP & Conv & Total \\ \hline
\multirow{11}{*}{ISCAS}											
&s1196	&	222	&	396	&	5	&	482	&	487	\\ \cline{2-7}
&s1238	&	245	&	385	&	5	&	487	&	492	\\ \cline{2-7}
&s1423	&	240	&	425	&	5	&	475	&	480	\\ \cline{2-7}
&s1488	&	307	&	395	&	5	&	474	&	479	\\ \cline{2-7}
&s5378	&	307	&	448	&	5	&	288	&	293	\\ \cline{2-7}
&s9234	&	304	&	422	&	5	&	204	&	209	\\ \cline{2-7}
&s13207	&	271	&	470	&	7	&	422	&	429	\\ \cline{2-7}
&s15850	&	253	&	248	&	7	&	324	&	331	\\\cline{2-7}
&s35932	&	358	&	667	&	7	&	576	&	583	\\ \cline{2-7}
&s38417	&	297	&	638	&	9	&	623	&	632	\\ \cline{2-7}
&s38584	&	204	&	656	&	8	&	346	&	354	\\ \hline
\multirow{4}{*}{CEP}											
& AES	&	1617	&	4974	&	13	&	11524	&	11537	\\ \cline{2-7}
& DES3	&	270	&	463	&	5	&	465	&	470	\\ \cline{2-7}
& SHA256	&	203	&	400	&	9	&	322	&	331	\\ \cline{2-7}
& MD5	&	420	&	495	&	12	&	527	&	539	\\ \hline
\multirow{3}{*}{CPU}											
&Plasma	&	226	&	549	&	23	&	70	&	93	\\ \cline{2-7}
& RISC-V	&	412	&	1037	&	17	&	1005	&	1022	\\ \cline{2-7}
&ARM-M0	&	379	&	925	&	29	&	485	&	514	\\ \hline
    \end{tabular}
    \caption{Run-times (sec) of our experiments}
    \label{tab:runtime}
\end{table}
The improvement over master-slave latch-based designs are more consistent and not as significant as FF-based designs. In particular, the maximal power deduction is 40\%, and an average of 12\%, 26\%, and 30\% benefit over ISCAS, CEP, and CPU master-slave designs. 
The overall power savings drop from 41\% to 16\% in the comparison changing from FF to master-slave designs. This can be explained by the fact that latch-based designs often have less glitching and fewer hold buffers than their FF-based counterparts.

Table \ref{tab:activity_power_results} reports the power dissipation of the resulting designs using switch-activity based power analysis assuming a switching activity of 20\% on all inputs (except reset and clocks) and registers. 
It shows similar savings as in the simulation-based power analysis shown in Table \ref{tab:power_results}. 

In summary, our experiments suggest that while significant saving in area and power is possible with our proposed approach, the amount of savings is variable and likely depends on a combination of factors including 1) the percentage of FFs with combinational feedback that limits the savings in number of latches and 2) the impact in retiming latch-based designs on the combinational logic. We should also note that these results are post synthesis and thus do not reflect the cost of the multiple clock trees nor the savings in hold buffers, both realized during physical design.

The run-time details of the conversion algorithm are reported in Table \ref{tab:runtime}. 
The column labeled ``FF Total" shows the run-times of FF-based synthesis, the next column corresponds to the run-time of master-slave latch-based design conversion, and the last three columns reports the run-times spent on solving ILP, converting and retiming, and the total for 3-phase latch-based designs.
Notice that the run-times for most designs, except for AES, are less than 18 minutes, in which at most 29 seconds is consumed by the ILP solver. 
This suggests that our proposed approach is computationally practical for at least moderately-sized blocks.
AES has the most number of registers (9703 FFs in the original design), and takes the longest time for conversion and retiming, i.e. 1 hrs 23 min for master-slave and 3 hrs 12 min for 3-phase.

\section{Conclusions}

\label{sec:conclusion}
This paper presents an algorithm to automatically convert a FF-based design into a 3-phase latch-based design 
that uses an ILP to minimize the number of required latches. 
Our experimental synthesis results on a broad range of benchmark circuits show significant savings are possible in both area and power with practical computational run-times, particularly for pipelined circuits such as multi-stage CPUs when compared to both FF and master-slave latch-based designs.

Our future work includes quantifying these benefits post place-and-route, including capturing the cost of routing multiple clock trees and 
the benefits associated with higher tolerance to PVT variations and increased robustness to hold failures. 
In addition, we plan to quantify the advantage of this approach when applied to timing and soft-error resilient templates in which the decrease in latches also reduces the overhead of the necessary error detection logic.

\vspace*{0.25cm}

\small{
\def\bibfont{\footnotesize}
\bibliographystyle{IEEEtran}
\vspace*{0.1cm}
\bibliography{output.bbl}

\begin{thebibliography}{10}
\providecommand{\url}[1]{#1}
\csname url@samestyle\endcsname
\providecommand{\newblock}{\relax}
\providecommand{\bibinfo}[2]{#2}
\providecommand{\BIBentrySTDinterwordspacing}{\spaceskip=0pt\relax}
\providecommand{\BIBentryALTinterwordstretchfactor}{4}
\providecommand{\BIBentryALTinterwordspacing}{\spaceskip=\fontdimen2\font plus
\BIBentryALTinterwordstretchfactor\fontdimen3\font minus
  \fontdimen4\font\relax}
\providecommand{\BIBforeignlanguage}[2]{{%
\expandafter\ifx\csname l@#1\endcsname\relax
\typeout{** WARNING: IEEEtran.bst: No hyphenation pattern has been}%
\typeout{** loaded for the language `#1'. Using the pattern for}%
\typeout{** the default language instead.}%
\else
\language=\csname l@#1\endcsname
\fi
#2}}
\providecommand{\BIBdecl}{\relax}
\BIBdecl

\bibitem{Haring2005}
R.~A. Haring, R.~Bellofatto, A.~A. Bright, P.~G. Crumley, M.~B. Dombrowa, S.~M.
  Douskey, M.~R. Ellavsky, B.~Gopalsamy, D.~Hoenicke, T.~A. Liebsch, J.~A.
  Marcella, and M.~Ohmacht, ``Blue gene/{L} compute chip: Control, test, and
  bring-up infrastructure,'' \emph{IBM Journal of Research and Development},
  vol.~49, no. 2.3, pp. 289--301, March 2005.

\bibitem{singh2018low}
K.~Singh, H.~Jiao, J.~Huisken, H.~Fatemi, and J.~P. De~Gyvez, ``Low power latch
  based design with smart retiming,'' in \emph{Quality Electronic Design
  (ISQED), International Symposium on}.\hskip 1em plus 0.5em minus 0.4em\relax
  IEEE, 2018, pp. 329--334.

\bibitem{Pons2016}
M.~Pons, T.~Le, C.~Arm, D.~Séverac, J.~Nagel, M.~Morgan, and S.~Emery,
  ``Sub-threshold latch-based icyflex2 32-bit processor with wide supply range
  operation,'' in \emph{2016 46th European Solid-State Device Research
  Conference (ESSDERC)}, Sept 2016, pp. 33--36.

\bibitem{Hurst2006}
A.~P~Hurst and R.~K~Brayton, ``The advantages of latch-based design under
  process variation,'' in \emph{Proceedings of the IWLS}, 2006.

\bibitem{fojtik2012bubble}
M.~Fojtik, D.~Fick, Y.~Kim, N.~Pinckney, D.~Harris, D.~Blaauw, and
  D.~Sylvester, ``Bubble razor: An architecture-independent approach to
  timing-error detection and correction,'' in \emph{Solid-State Circuits
  Conference Digest of Technical Papers (ISSCC), 2012 IEEE
  International}.\hskip 1em plus 0.5em minus 0.4em\relax IEEE, 2012, pp.
  488--490.

\bibitem{hand2015blade}
D.~Hand, M.~T. Moreira, H.-H. Huang, D.~Chen, F.~Butzke, Z.~Li, M.~Gibiluka,
  M.~Breuer, N.~L.~V. Calazans, and P.~A. Beerel, ``Blade--a timing violation
  resilient asynchronous template,'' in \emph{ASYNC}.\hskip 1em plus 0.5em
  minus 0.4em\relax IEEE, 2015, pp. 21--28.

\bibitem{lin2014low}
J.-F. Lin, ``Low-power pulse-triggered flip-flop design based on a signal
  feed-through scheme,'' \emph{IEEE Transaction on Very Large Scale Integration
  (VLSI) Systems}, vol.~22, no.~1, pp. 181--185, 2014.

\bibitem{paik2010statistical}
S.~Paik, L.-e. Yu, and Y.~Shin, ``Statistical time borrowing for pulsed-latch
  circuit designs,'' in \emph{Proceedings of the 2010 Asia and South Pacific
  Design Automation Conference}.\hskip 1em plus 0.5em minus 0.4em\relax IEEE
  Press, 2010, pp. 675--680.

\bibitem{singh2018multi}
K.~Singh, O.~A.~R. Rosas, H.~Jiao, J.~Huisken, and J.~P. de~Gyvez, ``Multi-bit
  pulsed-latch based low power synchronous circuit design,'' in \emph{Circuits
  and Systems (ISCAS), 2018 IEEE International Symposium on}.\hskip 1em plus
  0.5em minus 0.4em\relax IEEE, 2018, pp. 1--5.

\bibitem{Ding2017}
Y.~Ding, W.~Jin, G.~He, and W.~He, ``Short path padding with multiple-{Vtcells}
  for wide-pulsed-latch based circuits at ultra-low voltage,'' in \emph{2017
  IEEE 12th International Conference on ASIC (ASICON)}, Oct 2017, pp. 985--988.

\bibitem{shin2011pulsed}
Y.~Shin and S.~Paik, ``Pulsed-latch circuits: A new dimension in asic design,''
  \emph{IEEE Design \& Test of Computers}, vol.~28, no.~6, pp. 50--57, 2011.

\bibitem{yoshikawa2004timing}
K.~Yoshikawa, Y.~Hagihara, K.~Kanamaru, Y.~Nakamura, S.~Inui, and T.~Yoshimura,
  ``Timing optimization by replacing flip-flops to latches,'' in
  \emph{Proceedings of the Asia and South Pacific Design Automation
  Conference}.\hskip 1em plus 0.5em minus 0.4em\relax IEEE Press, 2004, pp.
  186--191.

\bibitem{cortadella2006desynchronization}
J.~Cortadella, A.~Kondratyev, L.~Lavagno, and C.~P. Sotiriou,
  ``Desynchronization: Synthesis of asynchronous circuits from synchronous
  specifications,'' \emph{IEEE Trans. on CAD}, vol.~25, no.~10, pp. 1904--1921,
  2006.

\bibitem{branover2004asynchronous}
A.~Branover, R.~Kol, and R.~Ginosar, ``Asynchronous design by conversion:
  Converting synchronous circuits into asynchronous ones,'' in
  \emph{Proceedings of the conference on Design, Automation and Test in
  Europe-Volume 2}.\hskip 1em plus 0.5em minus 0.4em\relax IEEE Computer
  Society, 2004, pp. 870--875.

\bibitem{Saifhashemi2014}
A.~Saifhashemi, D.~Hand, P.~A. Beerel, W.~Koven, and H.~Wang, ``Performance and
  area optimization of a bundled-data {Intel} processor through resynthesis,''
  in \emph{ASYNC}, May 2014, pp. 110--111.

\bibitem{zhang2018edge}
Y.~Zhang, H.~Cheng, D.~Chen, H.~Fu, S.~Agarwal, M.~Lin, and P.~A. Beerel,
  ``Challenges in building an open-source flow from {RTL} to bundled-data
  design,'' in \emph{Asynchronous Circuits and Systems (ASYNC), IEEE
  International Symposium on}, 2018.

\bibitem{cheng2018automatic}
H.~Cheng, H.-L. Wang, M.~Zhang, D.~Hand, and P.~A. Beerel, ``Automatic retiming
  of two-phase latch-based resilient circuits,'' \emph{IEEE Transactions on
  Computer-Aided Design of Integrated Circuits and Systems}, 2018.

\bibitem{sakallah1990optimal}
K.~A. Sakallah, T.~N. Mudge, and O.~A. Olukotun, ``Optimal clocking of
  synchronous systems,'' in \emph{In ACM International Workshop on Timing
  Issues in the Specification and Synthesis of Digital Systems}, 1990, pp.
  1--21.

\bibitem{ISCAS}
``{ISCAS89}: International symposium on circuits and systems sequential
  benchmark.
  {http://www.pld.ttu.ee/{\textasciitilde}maksim/benchmarks/iscas89/verilog/}.''

\bibitem{CEP}
``{MIT-LL} common evaluation platform ({CEP}),''
  \url{https://github.com/mit-ll/CEP}, available: 2019.

\bibitem{plasma}
``Plasma {CPU},'' \url{http://opencores.org/project,plasma}, available: 2014.

\bibitem{rocket}
``Rocket chip,'' \url{https://github.com/freechipsproject/rocket-chip},
  available: 2016.

\bibitem{armm0}
``{ARM} {C}ortex {M0},''
  \url{https://developer.arm.com/products/processors/cortex-m/cortex-m0}.

\bibitem{singhal1997case}
V.~Singhal, S.~Malik, and R.~K. Brayton, ``The case for retiming with explicit
  reset circuitry,'' in \emph{Proceedings of the 1996 IEEE/ACM international
  conference on Computer-aided design}.\hskip 1em plus 0.5em minus 0.4em\relax
  IEEE Computer Society, 1997, pp. 618--625.

\bibitem{gurobi}
\BIBentryALTinterwordspacing
L.~Gurobi~Optimization, ``Gurobi optimizer reference manual,'' 2018. [Online].
  Available: \url{http://www.gurobi.com}
\BIBentrySTDinterwordspacing

\bibitem{chinnery2002closing}
D.~Chinnery and K.~Keutzer, \emph{Closing the gap between ASIC \& custom: tools
  and techniques for high-performance ASIC design}.\hskip 1em plus 0.5em minus
  0.4em\relax Springer Science \& Business Media, 2002.

\bibitem{brglez1989combinational}
F.~Brglez, D.~Bryan, and K.~Kozminski, ``Combinational profiles of sequential
  benchmark circuits,'' in \emph{IEEE International Symposium on Circuits and
  Systems}, 1989, pp. 1929--1934.

\end{thebibliography}
}
\end{document}